\documentstyle[12pt]{article}
\begin{document}
\def\theequation{\arabic{section}.\arabic{equation}}
\newcommand{\beq}{\begin{equation}}
\newcommand{\eeq}{\end{equation}}
\title {Hamiltonian Structures on Coadjoint Orbits of
 Semidirect Product $G=Diff_+(S^{1})\triangleright
 \! \! \!{<}{C}^{\infty}(S^{1}, {R}{\! \! \! \! \! \,}{R})$ }
\bigskip
\author{ \bf A. Zujewski \thanks{email: zuevsky@vxitep.itep.ru }
\bigskip \\
\centerline{\em Moscow Institute of Physics and Technology}\\
\centerline{\em and}\\
\centerline{\em Institute of Theoretical and Experimental Physics,}\\
\centerline{\em B.Cheremushkinskaya 25, 117259,}\\
\centerline{\em Moscow, Russia}\\
\centerline{\em hepth/eeeeddd}
}
\maketitle
\begin{abstract}

 We consider the semidirect product of diffeomorphisms of the circle
$D={Diff}_+(S^1)$ and $C^{\infty}(S^{1}, {R}{\! \! \! \! \! \,}{R})$
functions, classify its coadjoint orbits and treat dynamical
systems related to corresponding Lie algebra centrally extended by
Kac-Moody, Virasoro and semidirect product cocycles with arbitrary
coefficients.

 The isomorphism (under certain conditions on elements of the
coalgebra in the form of a Miura transformation) between Lax $L$-operators
and smooth orientation-preserving immersions $S^1 \longrightarrow {{P}{\! \!
\! \! \!  \,}{P}}^1 \times {{P}{\! \! \! \! \!  \,}{P}}^1 $ is established .

 Using the three-hamiltonian structure formalism we prove the integrability
of our generalized system in the case of a family of Hamiltonians
formed from residues of powers of pseudo-differential operators.
Relations between DWW and mKdV hierarchies appear when we introduce an
alternative sequence of Hamiltonians.

 The construction of commuting flows on coadjoint orbits is
generalized for our group and Poisson bracket in the coalgebra of the
associated affine loop algebra is presented.

\end{abstract}
\newpage


\section{Introduction: Central extensions of semidirect product
${\cal G}=Vect(S^1)\triangleright \! \! \!{<}{C}^{\infty}(S^{1},
{R}{\! \! \! \! \! \,}{R})$ }

\qquad{}  In the series of papers the Lie algebra of the group of
orientation preserving diffeomorphysms of the circle ${\cal
D}=Diff_+(S^{1})$ was considered (see \cite{kirillov}, \cite{witten},
\cite{segal1}). Authors of \cite{marshall} and \cite{kupershmidt1} treat
 the semidirect product of ${\cal D}$ and the space of smooth non-vanishing
functions ${\cal V}= C^{\infty}(S^{1},{R}{\! \! \! \! \!  \,}{R})$ on
$S^{1}$ with central extensions by three cocycles $C_{K-M}, C_{SP}$ and
 $C_{Vir}$ of corresponding algebra of differential operators introduced in
\cite{arbarello}.

  In \cite{kirillov} and \cite{witten} one can find a classification of
coadjoint orbits of group ${\cal D}=Diff_+(S^{1})$ the algebra of which is
centrally extended by the Virasoro cocycle.

 The Lax representation for the dynamical system occurred from the coadjoint
action of the extended by three cocycles algebra of the above mentioned
semidirect product was found in \cite{marshall}. The theorem in
\cite{segal1} established an isomorphysm between the space of Hill equations
and the space of smooth orientation-preserving immersions of $S^1
\longrightarrow {{P}{\! \! \! \! \! \,}{P}}^1$, (${{P}{\! \! \! \! \!
 \,}{P}}^1$ is the real projective line). Different relations to the KdV
hierarchy and procedure of the construction of commuting flows on the phase
space are also presented there.

  The author of \cite{kupershmidt2} has developed the theory of the
non-standard integrable systems and applied to a particular case of the
operator ${L}{\!  \! \!  \! \,}{I}=b\partial^{-1}+a+\beta\partial$.
 The existence of the three-hamiltonian structure form of the DWW (Dispersive
Water Waves) system of equations and its integrability were proved due to
 properties of ${L}{\!  \! \!  \! \,}{I}$ operators. The second hamiltonian
structure is a matrix coadjoint representation of the Lie algebra with a
particular case of the linear combination of two cocycles in the central
extension of the algebra.

The main purpose of this paper is to show that the dynamical systems
that occur from the coadjoint action of the algebra ${\cal
G}=Vect(S^1)\triangleright \! \! \!{<}{C}^{\infty}(S^{1}, {R}{\! \! \! \!
\!  \,}{R})$ centrally extended by three cocycles with arbitrary
coefficients are also integrable.

The plan of the paper is the following. Section 1 describes our algebra and
its central extensions.

In Section 2 we show that the classification of its coadjoint orbits is
closely related to the classification in the case of the Virasoro algebra.
Namely the space of elements conserving the element of the coalgebra
is just the same as for the Virasoro algebra.

In Section 3 the theorem proved in \cite{segal1} is generalized. It turns
out that the space of the Lax operators is isomorphic to the space of
immersions $S^1 \longrightarrow {{P}{\! \! \!  \!  \!  \,}{P}}^1 \times
{{P}{\!  \! \!  \!  \!  \,}{P}}^1$ under certain restrictions on elements of
the coalgebra.  Those restrictions are related to the proof of the
integrability of our system.

  In Section 4 we present a commuting family of Hamiltonians,
corresponding to our algebra which gives us a generalization of the DWW
system.  The alternative sequence of Hamiltonians starting from the standard
quadratic Hamiltonian results in the KdV-like system of equations. In that
case we find two reductions to the pair of KdV and mKdV equations.

  In Section 5 we generalize the recipe of the construction of commuting
flows on the phase space (see \cite{segal1}) for the centrally extended
algebra.

   The Poisson bracket can be introduced for the affine loop algebra analog of
the Virasoro algebra \cite{semenov2}. In the end of this paper we present
the first and the second systems of compatible Poisson
brackets for DWW and KdV-like systems corresponding to the extended algebra
of semidirect product $G={\cal D}\triangleright \! \! \!{<}{\cal V}$.

  As for physical application, we show that the KdV-like pair of equations
describes a system of generalized magnetic hydrodynamics on the phase space
of our group \cite{vishik}. We also consider small oscillations of our
system.


\bigskip
\bigskip


\qquad {} Let ${\cal D}=Diff_+(S^{1})$ be the group of diffeomorphisms of
 $S^{1}$ and ${\cal V} =C^{\infty}(S^{1},{R}{\! \! \! \! \! \,}{R})$ the
space of smooth non-vanishing functions on $S^{1}$. Consider a semidirect
product $G=\cal D \triangleright \! \! \!{<}\cal V$ with the action of $\cal
D$ on  $\cal V$ (see \cite{marshall})

\beq
\label{actionalso}
 {\varrho }\;\nu=\nu \circ {\varrho}
\eeq

where ${\varrho} \in \cal D $ and $\nu \in \cal V $.
The group multiplication in $G$ is

\beq
\label{multigroup}
({\varrho}_{1},{\nu}_{1})({\varrho}_{2},{\nu}_{2})=({\varrho }_{2}
  \circ {\varrho}_{1},\;{\nu}_{1}+{\nu}_{2} \circ {\varrho }_{1}).
\eeq
\bigskip

 The Lie algebra corresponding to $G$ is the semidirect product
${\cal G}=Vect\; (S^1) \triangleright \! \!{<} {\cal V}$ of the algebra of
vector fields $Vect\; (S^1)$ on $S^{1}$ and ${\cal V}
=C^{\infty}(S^{1},{R}{\!  \! \! \! \!  \,}{R})$. We can consider a central
extension \\ $\widehat{\cal G} = \cal {G}$ $\oplus$ $ {R}{\! \! \! \! \!
\,}{R}$ $ \oplus$ $ {R}{\! \! \!  \! \!  \,}{R}$ $\oplus $ ${R}{\!  \! \! \!
  \! \,}{R} $ of ${\cal G}$ by the following cocycles (\cite{marshall},
  \cite{arbarello})

\beq
C_{K-M} (( f_{1} , g_{1}) ,(f_{2},g_{2}))=\int_{S^1} g_{1} g_{2}'\; dx
\label{CK-M}
\eeq

\beq
C_{SP}((f_{1},g_{1}),(f_{2},g_{2}))=\int_{S^1}( f_{1}'' g_{2}-f_{2}''g_{1})\;
dx
\label{Cpp}
\eeq

\beq
C_{Vir} ( ( f_{1} , g_{1} ),(f_{2} , g_{2}))=\int_{S^1}f_{1}f_{2}'''\; dx
\label{CVir}
\eeq
\bigskip

( primes mean $\partial =\frac {\partial} {\partial x}$ where $x$ is a
coordinate on $S^{1}$).

The Lie algebra structure in $\widehat {\cal G}$ is given by the bracket

\beq
\begin{array}{l} \bigskip
\left[ \; (f_{1},g_{1},{({\alpha}_{1},{\alpha}_{2},{\alpha}_{3})}_{1}),
(f_{2},g_{2},{({\alpha}_{1},{\alpha}_{2},{\alpha}_{3})}_{2}) \;  \right] =\\
\qquad {}\qquad {} (f_1f_2'-f_2f_1',\; f_1g_2'-f_2g_1' ,\;C_{K-M},\;C_{SP},
\;C_{Vir}).
\end{array}
\label{commutator}
\eeq

 We associate the dual space ${\widehat {\cal G}}^{*}$ ( the space of
pseudo-differential operators on $S^{1}$) for ${\widehat {\cal {G}}} $ by
 means of the pairing between $(f,g,{\alpha}_{1},{\alpha}_{2},{\alpha}_{3})
 \in  \widehat{\cal{G}}$ and $(b ,a ,{\beta}_{1},{\beta}_{2},{\beta}_{3} )
 \in {\widehat{\cal{G}}}^{*}$

\beq
\langle (f,g,{\alpha}_{1},{\alpha}_{2},{\alpha}_{3} ),
(b, a ,\beta_{1},\beta_{2},\beta_{3})\rangle =
\int_{S^1} (fb+ga)\;dx+\sum\limits_{i=1}^{3}{\alpha}_{i}{\beta}_{i}.
\label{pairing}
\eeq

The coadjoint action of $\widehat {\cal G}$ on ${\widehat {\cal G}}^{*}$
 is the generalization of the coadjoint action of ${\cal G}$ on ${\cal
 G}^{*}$

\beq
\begin{array}{l}\bigskip
{\widehat{ad}}^*_{(f,g,{\alpha}_{1},{\alpha}_{2},{\alpha}_{3})}
(b,a,{\beta}_{1},{\beta}_{2},{\beta}_{3})=
(2f'b+fb'+ag'+{\beta}_{2}g''+{\beta}_{3}f''',\\
\qquad{} \qquad {} (fa)'+{\beta}_{1}g'-{\beta}_{2}f'',\;0,\;0,\;0).
\label{adzve}
\end{array}
\eeq

 We can also consider the extension ${\widehat {\cal G}}^{'}={\cal G} \oplus
{R}{\! \! \! \! \! \,}{R}$ of ${\cal G}$ by a linear combination

\beq \label{combination}
Z=l_{1}C_{K-M}+l_{2}C_{P-P}+l_{3}C_{Vir}
\eeq

(i.e. $\left\{f\partial+g+\alpha Z \right\} \in {\widehat {\cal G}}^{'}$)
with the appropriate Lie bracket

\beq
\label{otherbracket}
\left[ (f_1,g_1,\alpha_1),(f_2,g_2,\alpha_2) \right]=
(f_1f_2'-f_2f_1',\;f_1g_2'-f_2g_1',\;Z).
\eeq

 The pairing between $(f,g,\alpha) \in {\widehat {\cal G}}^{'}$
and $(b,a,\beta) \in {\widehat {\cal G}}^{'*}$
(i.e. $\left\{ b {\partial}^{-2}+ a {\partial}^{-1}+ \beta \log \partial
\right\} \in {\widehat {\cal G}}^{'*}$, \cite{wrubtzov}) of the form

\beq
\label{otherpairing}
\langle (f,g,\alpha),(b,a,\beta) \rangle =\int_{S^{1}}(fb+ag)\;dx+
\alpha \beta.
\eeq

and the coadjoint action

\beq
\label{otheraction}
\begin{array}{l}\bigskip
{{\widehat {ad'}}^{*}}_{(f,g,\alpha)} (b,a,\beta)=
(2f'b+fb'+ag'+ \beta (l_{2}g''+l_{3}f'''),\\
\qquad{} \qquad{} (fa)'+\beta (l_{1}g'-l_{2}f''),0).
\end{array}
\eeq

The meaning of a special linear combination with

\beq
l_{1}=1,\;l_{2}=-\frac{1}{2},\;l_{3}=\frac {1}{6}
\eeq

is explained in \cite{arbarello}.


\section{Coadjoint orbits of $G$ }

\qquad {} The orbit method due to Kirillov \cite{kirillov} can be applied to
the classification of coadjoint orbits of $G$. Let us determine the subgroup
$G_{\delta}$ of $G$ which leaves
 $\rho=(b,a,\beta_1,\beta_2,\beta_3) \in {\widehat {{\cal {G}}^*}}$ invariant
under the action of an element $\kappa=(f,g,\alpha_1,\alpha_2,\alpha_3) \in
{\widehat {\cal G}}$.  The transformation laws of $\rho$ under an
infinitesimal transformation by $\kappa$ are

\beq
\delta b=2f'b+fb'+ag'+\beta_{2}g''+\beta_{3}f'''=0,
\label{bvariacia}
\eeq

\beq
\delta a=(fa)'+\beta_{1}g'-\beta_{2}f''=0,
\label{avariacia}
\eeq

\beq
\delta \beta_1=\delta \beta_2=\delta \beta_3=0.
\label{betavariacia}
\eeq

Eliminating $g$ from (\ref{bvariacia}) with the help of (\ref{avariacia}) we
 get (for $\beta_{1} \ne 0$) a system

\beq
\left\{
\begin{array}{rcl}
\delta {\widetilde b}&=&\gamma f'''+2f'{\widetilde b}+f{\widetilde b}'=0 \\
\delta a&=&(fa)'+\beta_1g'-\beta_2f''=0\\
\delta \beta_1&=&\delta \beta_2=\delta \gamma =0
\end{array}
\right.
\label{fbtilda}
\eeq

where

\beq
\label{btilda}
\begin{array}{c}
\bigskip
 {\widetilde b}=
b-\frac {a^{2}} {2\beta_{1}}-\frac {\beta_{2}} {\beta_{1}} a'\\
\gamma=\frac {{\beta_2}^2}{\beta_1} +\beta_3.
\end{array}
\eeq

 The first equation in (\ref{fbtilda}) is a condition for $\widetilde {b}$
to be invariant under the action of the $f$  which is an
element of the Virasoro algebra \cite{kirillov}. Therefore we have just the
same classification of $f$-elements and ${\widetilde b}$ as in
\cite{kirillov},\cite{witten}.

  If $\frac{{\beta_2}^2}{\beta_1} +\beta_3=0, \beta_1 \ne 0$ then $f$ is an
element of non-extended algebra of vector fields \cite{witten} (see also
\cite{marshall}).

 The classification of coadjoint orbits for the $g$-element of ${\widehat
{\cal G}}$ can be deduced from the second equation of (\ref{fbtilda}).
Namely

\beq
\label{aclass}
\beta_1g=const-\beta_2f'-fa.
\eeq

 We see that the $g$-element of ${\widehat {\cal G}}$ is related to the
$f$-element by means of (\ref{aclass}) with the functional parameter $a$.

  In principle, one can eliminate from (\ref{bvariacia}) the element
corresponding to the cocycle $C_{Vir}$. Let us differentiate
(\ref{avariacia}) multiplied by $\frac{\beta_3}{\beta_2}, \;(\beta_2\ne0)$,
add (\ref{avariacia}) multiplied by $\frac{a\beta_3}{{\beta_2}^2}$
and subtract the sum from (\ref{bvariacia}) then

\beq
\label{fbstar}
\left\{
\begin{array}{rcl}
\delta b_*&=& 2f'b_*+fb_*'+ag'+\beta_{2}g''=0\\
\delta a&=& (fa)'+\beta_{1}g'-\beta_{2}f''=0\\
\delta \beta_1&=&\delta \beta_2 =0
\end{array}
\right.
\eeq

where

\beq
\label{bstar}
b_*=\frac{1}{1+\frac{\beta_3\beta_1}{{\beta_2}^2}}
  (b+\frac{\beta_3}{\beta_2}a'+\frac{\beta_3}{2{\beta_2}^2}a^2).
\eeq

  Let us show that we can transform the $g$-component of the algebra in such
a way that the element of ${\widehat{\cal G}}^*$ corresponding to $C_{PP}$
will not appear explicitly in the coadjoint action. Making the substitution

\beq
\label{breve g}
{\breve g}=\frac{\beta_1}{2\beta_2}g
\eeq

we can rewrite the system (\ref{fbstar}) as

\beq
\label{euforustilde}
\left\{
\begin{array}{rcl}
\delta {\breve b}_* &=& 2f'{\breve b}_*+f{{\breve b}_*}'+a{\breve
g}' +\beta_2 {\breve g}''=0 \\
\delta a &=& (af)'+2\beta_2 {\breve
g}'-\beta_2f''=0\\
\delta \beta_2&=&0
\end{array} \right.
\eeq

with

\beq
\label{bbreve}
{\breve b}_*=\frac{\beta_1}{2\beta_2}b_*.
\eeq

 The consequences of such a transformation will be clear in Section 4.


\section{Lax operator representation for elements of ${\widehat {\cal
{G}}}$ and ${\widehat {\cal {G}}}^{*}$ }

\qquad The element $V \in {\cal G}$ may be represented as a matrix
(see \cite{marshall})

\beq
\label{diffoper}
V={\left(
\begin{array}{ccc}
 -\frac {1}{2} f'+f \circ \partial &{}& 0 \\
 {}&{}&{}\\
-g' &{}& \frac {1}{2} f'+f \circ \partial
\end{array} \right)}_.
\eeq

 The set of such differential operators forms a Lie algebra under ordinary
commutator of operators. We can introduce the action of $V$ on $L
\in {\widehat {\cal G}}^*$

\beq
\label{diffoperaction}
 {ad}^*_V L = V \circ L=[V,L]+AL
\eeq

where

\beq
\label{amatrix}
A=\left( \begin{array}{cc}
2f' & -g'\\g' & 0 \end{array} \right)
\eeq

and

\beq
\label{ourL}
L=\left(
\begin{array}{ccc}
2b+4\beta_{3} \circ  {\partial}^{2} &{}& 2\beta_{2} \circ \partial +a\\
{}&{}&{}\\
2\beta_{2} \circ \partial -a &{}& -\beta_{1}\\
\end{array} \right)
\eeq

which is the coadjoint action in that representation. Then the Lax form of
the (\ref{bvariacia}),(\ref{avariacia}) and (\ref{betavariacia}) is

\beq
\label{laxform}
\delta L= {ad}^*_V L=0.
\eeq

 The modification of $L$-operator for the system (\ref{fbstar}) is
transparent

\beq
\label{ourLstar}
L_*={\left(
\begin{array}{ccc}
2b_* &{}& 2\beta_{2} \circ \partial +a\\
{}&{}&{}\\
2\beta_{2} \circ \partial -a &{}& -\beta_{1}\\
\end{array} \right)}_.
\eeq

 Let $\psi$ be a column vector of two variables

\beq
\label{phi}
\psi={f_1 \choose g_1}.
\eeq

 Consider the linear problem

\beq
\label{lphi}
L \psi =0
\eeq

which is equivalent to

\beq
\label{lpsisys}
\left\{ \begin{array}{rcl}
2bf_1+4\beta_{3}f_1''+2\beta_{2}g_1'+ag_1 & = & 0\\
2\beta_{2}f_1'-af_1-\beta_{1}g_1 & = & 0.\\
\end{array} \right.
\eeq

 Eliminating $g_1$ from the first equation of (\ref{lpsisys}) we arrive at
Hill equation (\cite{marshall},\cite{lazutkin})

\beq
\label{hill}
f_1''+\frac {1} {2\gamma}\; \widetilde {b}f_1=0.
\eeq

  Let us take two independent solution  $(f_1,g_1)$ and $(f_2,g_2)$ of
(\ref{lpsisys}). Then the second equation of it ( for $\beta_1 \ne 0$ )
gives

\beq
\label{gqua}
 g_1=\frac{1}{\beta_1}(af_1-2\beta_2f_1').
\eeq

It easy to check that the Wronskian

\beq
\label{wwronskians}
\begin{array}{c}
\bigskip
 W(g_1,g_2)=\frac{1}{{\beta_1}^2}
\left(a^2+2\beta_2a'+2\frac{{\beta_2}^2}{\gamma}{\widetilde b}\right)
W(f_1,f_2) \\
\bigskip
= \frac{2}{\beta_2}\left(b+\frac{\beta_3\beta_1} {\beta_1\beta_3+{\beta_2}^2}
{\widetilde b}\right)W(f_1,f_2).
\end{array}
\eeq

 For the operator (\ref{ourLstar}) using $L_*$ we get the following condition

\beq
\label{wwronskianstar}
 W_*(g_1,g_2)=\frac{1}{{\beta_1}^2}b_*W_*(f_1,f_2).
\eeq

Since the Wronskian $ W_*(f_1,f_2)=C_0 $ is a constant, therefore
the space of $L_*$ operators is isomorphic to the
space of smooth orientation preserving immersions $S^1 \longrightarrow
{{P}{\! \! \! \! \! \,}{P}}^1 \times {{P}{\! \! \! \! \! \,}{P}}^1$ when

\beq
\label{bstarcond}
b_* = const(x)
\eeq
(compare with \cite{segal1}). One can mention that (\ref{bstarcond}) is some
sort of Miura transformation as well as (\ref{btilda}),(\ref{bstar}). We can
only repeat that the same situation takes place for the space of
$L$-operators when

\beq
b+\frac{\beta_3}{\gamma}{\widetilde b} = const(x).
\eeq


\section{Hamiltonian structures on coadjoint orbits of $G$ }

 On coadjoint orbits of a group we could define as in \cite{segal1} a
classical mechanics system by the following Euler equation for $\chi \in
{\widehat {\cal {G}}}^*$ (dot means the coadjoint action) (see
\cite{arnold})

\beq
{\partial}_t \chi =({M}^{-1}\chi).\chi
\label{eulerequation}
\eeq

where the product in the r.h.s. is the coadjoint action of
$\widehat {\cal G}$ on ${\widehat {\cal G}}^{*}$.
(In the case of ${SO}_{3}$, $\chi$ and $M$ are the angular momentum and the
inertia tensor of free moving body in the configuration space $SO_3$
correspondingly).

  The Euler equation (\ref{eulerequation}) for the case of $G$ may be
written in the form

\beq
\label{abm}
{{a \choose b}}_t={\left( M^{-1}{a \choose b} \right)}.{a \choose b}
\eeq

or as a coadjoint action

\beq
\label{euforus}
\left\{ \begin{array}{rcl}
{\dot b} &=& 2f'b+fb'+ag'+\beta_{2}g''+\beta_{3}f'''\\
{\dot a} &=&(fa)'+\beta_{1}g'-\beta_{2}f''.\\
\end{array} \right.
\eeq

(dots mean $\partial_t$). Using methods of Section 3 (equations
(\ref{fbtilda}), (\ref{fbstar})) we obtain two alternative formulations of
the dynamical system (\ref{euforus})

\beq
\label{btildadin}
\left\{
\begin{array}{rcl}
\dot {\widetilde b}&=&\gamma f'''+2f'{\widetilde b}+f{\widetilde b}'\\
\dot a&=&(fa)'+\beta_1g'-\beta_2f''
\end{array}
\right.
\eeq

and

\beq
\label{bstardin}
\left\{ \begin{array}{rcl}
{\dot b_*} &=& 2f'b_*+fb_*'+ag'+\beta_{2}g''\\
{\dot a} &=&(fa)'+\beta_{1}g'-\beta_{2}f''.\\
\end{array} \right.
\eeq

 We may write (\ref{euforus}) and (\ref{bstardin}) in a hamiltonian form

\beq
{ {a\choose b} }_t =
   B {\frac {\delta {H}}{\delta a} \choose
\frac {\delta {H}}{\delta b}}
\label{eulerkuper}
\eeq

 with the column vector of variational derivatives of a Hamiltonian $H$ in
the r.h.s. and $B$ as a hamiltonian matrix (see \cite{kupershmidt2}).
Identifying elements of our algebra with elements of the coalgebra ( i.e.
linear parts of variations of $H$ by $a$ and $b$) let us take $f=a$ and
$g=b$ for (\ref{euforus}) or $f=a$ and $g=b_*$ for (\ref{bstardin}). Then we
find that in (\ref{eulerequation})

\beq
\label{mhsecond}
M^{-1}=M^{-1}_{DWW}=
\left(
\begin{array}{cc}
0&1\\
1&0
\end{array}
\right)
\eeq

and arrive at two generalizations of the DWW system  of
equations (\cite{kupershmidt2})

\beq
\label{mdww}
\left\{
\begin{array}{rcl}
{\dot  a} &=&(\beta_{1}b+a^2-\beta_{2}a')'\\
{\dot b} &=&(2ab+\beta_{2}b'+\beta_{3} a'')'
\end{array}
\right.
\eeq

and

\beq
\label{bstarkudin}
\left\{
\begin{array}{rcl}
{\dot  a} &=&  (\beta_{1}b_*+{a}^{2}-\beta_{2}a')'\\
{\dot b_*} &=&  (2ab_*+\beta_{2}b_*')'.
\end{array}
\right.
\eeq

 A substitution (see (\ref{bbreve}))

\beq
\label{bbreve1}
{\breve b}_*=\frac{\beta_1}{2\beta_2}b_*
\eeq

allows us to rewrite the system (\ref{bstarkudin}) as

\beq
\label{bstarkudinbreve}
\left\{
\begin{array}{rcl}
{\dot  a} &=&  (2\beta_{2}b_*+{a}^{2}-\beta_{2}a')'\\
{\dot {\breve b}_*} &=&  (2a{\breve b}_*+\beta_{2}{\breve b}_*')'
\end{array}
\right.
\eeq

and in the form (\ref{eulerkuper}) with the matrix $B={\breve B}^{II}_*$

\beq
\label{biistarbreve}
{\breve B}^{II}_* = \left(
\begin{array}{ccc}
2\beta_2 \circ \partial && \partial
 \circ a - \beta_{2} \circ {\partial}^2 \\
 &&
 \\ a \circ \partial + \beta_{2}
 \circ {\partial}^2 && {\breve b}_* \circ \partial + \partial \circ {\breve
b}_*
\end{array}
\right)
\eeq

( substitutions $b_* \mapsto {\breve b}_*$ should be made in the column
vector of variational derivatives in (\ref{eulerkuper})).

  One can mention that (\ref{bstarkudinbreve}) is the DWW system of
equations \cite{kupershmidt1}. It was proved in \cite{kupershmidt2} that
such a system is integrable and there exist the following three hamiltonian
structure formulation for it ( ${\breve B}^{II}_*$ coincides with the second
hamiltonian structure)

\beq
\label{brevetthreestru}
{ {a \choose {\breve b}_*} }_t =
{\breve B}^{I}_*
\left(
\begin{array}{c}
\frac {\delta {\breve H}_{k+1} } {\delta a}\\
{}\\
\frac {\delta {\breve H}_{k+1} } {\delta {\breve b}_*}
\end{array}
\right) =
{\breve B}^{II}_*
\left(
\begin{array}{c}
\frac {\delta {\breve H}_k} {\delta a}\\
{}\\
\frac {\delta {\breve H}_k} {\delta {\breve b}_*}
\end{array}
\right) =
{\breve B}^{III}_*
\left(
\begin{array}{c}
\frac {\delta {\breve H}_{k-1} } {\delta a}\\
{}\\
\frac {\delta {\breve H}_{k-1} } {\delta {\breve b}_*}
\end{array}
\right)
\eeq

where $k=1,2,...$. The first and third hamiltonian structures are given by

\bigskip
\beq
\label{kufirststrabreve}
{\breve
B}^{I}_*= \left( \begin{array}{cc} 0 & \partial \\
\partial & 0 \\
\end{array}
\right)
\eeq

and

\bigskip

\beq
\label{brevethirdhamstra}
{\breve B}^{III}_*= \! {\left( \!
\begin{array}{ll}
2\beta_2(\partial\circ a+a\circ\partial) &
\begin{array}{l}
  2\beta_2 (\partial \circ {\breve b}_*+{\breve b}_*\circ \partial)\\
  \qquad{} +\partial \circ {(a-\beta_2 \circ \partial)}^2\\
 \qquad{} \qquad{}
\end{array}
  \\
 {}&{}\\
\begin{array}{l}
2\beta_2(\partial\circ {\breve b}_*+{\breve b}_*\circ\partial)\\
\qquad{} -{(a+\beta_2\circ\partial)}^2\circ \partial \\
\qquad{} \qquad{}
\end{array}
 &
 \begin{array}{l}
(a+\beta_2\partial)({\breve b}_*\partial+\partial {\breve b}_*)\\
\qquad{} +({\breve b}_*\partial+\partial {\breve b}_*)
(a-\beta_2\partial)\\
\qquad{} \qquad{}
 \end{array}
\end{array}
\! \! \right)}_.
\eeq
\bigskip

The identity

\beq
\label{identl}
 {({{L}{\! \! \! \! \,}{I}}^\dagger)}^{m+1}=({{{L}{\! \! \! \! \,}{I}}
 ^\dagger})^m{{L}{\! \! \! \! \,}{I}}^\dagger={{L}{\! \! \! \! \,}{I}}
 ^\dagger{({{L}{\! \! \! \! \,}{I}}^\dagger)}^m
 \eeq

with the ${L}{\! \! \! \! \,}{I}$-operator ( ${ }^\dagger$ means adjoint)

\beq
\label{operatorl}
{L}{\! \! \! \! \,}{I}={\breve b}_*\partial^{-1}+a+\beta_2\partial
\eeq

was used in \cite{kupershmidt2} to prove (\ref{brevetthreestru}) for
the family of conserved commuting Hamiltonians (see also
\cite{kupershmidt1})

\beq
 {\breve H}_{n}=\frac{1}{n}\; Res \; {{L}{\! \! \! \! \,}{I}}^n=
 \frac{1}{n}\; Res \;
 {({\breve b}_*{\partial}^{-1}+a+ \beta_2\partial)}^{n}.
 \label{brevehamiltonians}
 \eeq

 If we expand ${({L}{\! \! \! \! \,}{I})}^m$ in $\xi$ ( $\xi$
denotes partial derivative)

\beq
\label{ryad}
{({L}{\! \! \! \! \,}{I})}^m=\sum_s {\breve p}_s(m)\xi^s
\eeq

and pick out the ${\xi}^i$ terms in (\ref{identl}) for $i=0,1,2$ then we
get \cite{kupershmidt2} for all $m$

\beq
\label{p0m1}
{\breve p}_0(m+1)=2\beta_2{\breve p}_{-1}(m)+
(a-\beta_2\partial){\breve p}_0(m)
\eeq
\beq
\label{p-1m1}
\partial {\breve p}_{-1}(m+1)=(a\partial+\partial^2){\breve p}_{-1}(m)+
({\breve b}_*\partial+\partial {\breve b}_* )p_0(m).
\eeq

   In the theory of non-standard integrable systems \cite{kupershmidt2} the
following relations occur

\beq
\label{p0}
{\breve p}_0(m)=\frac{\delta {\breve H}_{m+1} }{\delta {\breve b}_*}
\eeq
\beq
\label{p-1}
{\breve p}_{-1}(m)=\frac{\delta {\breve H}_{m+1} }{\delta a}
\eeq

which combined with (\ref{p0m1}) and (\ref{p-1m1}) gives the first and
second hamiltonian structures). Let us change variables

\beq
\label{tildep0m}
 p_0(m)={\breve p}_0(m)
\eeq
\beq
\label{tildep-1m}
p_{-1}(m)=\frac{2\beta_2}{\beta_1}{\breve p}_{-1}(m)
\eeq

so that

\bigskip
\beq
\label{newvarp0}
 p_0(m)=\frac{\delta H_{*m+1} }{\delta b_*}=
\frac{\delta {\breve H}_{m+1} }{\delta {\breve b}_*}
\eeq
\beq
\label{newvarp-1}
 p_{-1}(m)=\frac{\delta H_{*m+1} }{\delta a}=
\frac{2\beta_2}{\beta_1} \frac{\delta {\breve H}_{m+1} }{\delta a}
\eeq
\bigskip

and we can rewrite (\ref{p0m1}) and (\ref{p-1m1}) as

\beq
\label{ptilde0m1}
 p_0(m+1)=\beta_1 p_{-1}(m)+(a-\beta_2\partial) p_0(m)
\eeq
\beq
\label{ptilde1m1}
\partial p_{-1}(m+1)=(a\partial+\partial^2)p_{-1}(m)+
( b_*\partial+ \partial b_*) p_0(m)
\eeq

for all $m$. (\ref{ptilde0m1}) and (\ref{ptilde1m1}) give us the second
hamiltonian structure for systems (\ref{bstarkudin})

\bigskip
\beq
\label{biistar}
 B^{II}_* =
 {\left(
 \begin{array}{ccc}
\beta_{1} \circ \partial && \partial
 \circ a - \beta_{2} \circ {\partial}^2 \\
 &&
 \\ a \circ \partial + \beta_{2}
 \circ {\partial}^2 && b_* \circ \partial + \partial \circ b_*
 \end{array} \right)}_.
\eeq
\bigskip

  We derive from (\ref{newvarp0}) and (\ref{newvarp-1}) the family of
Hamiltonians

\beq
 H_{*n}= \frac{2\beta_2}{\beta_1n} \; Res \;
 {\left(\frac{\beta_1}{2\beta_2} b_*{\partial}^{-1}+a+
 \beta_2\partial\right)}^{n}
\label{kuhamiltonians}
 \eeq

and calculate the third hamiltonian structure using (\ref{ptilde0m1}) and
(\ref{ptilde1m1}) ( the first structure is just the same as in
(\ref{brevetthreestru}))

\bigskip

\beq
\label{thirdhamstrastar}
\! \!  B^{III}_*=
\! \! \! {\left(
\! \! \! \begin{array}{ll}
\beta_1 (\partial \circ a+a\circ \partial)
&
\begin{array}{l}
\beta_1 (\partial \circ b_*+b_*\circ \partial)+ \\
\qquad{} \partial \circ {(a-\beta_2 \circ \partial)}^2
  \end{array}
  \\
{}&{}\\
\begin{array}{l}
\beta_1 (\partial \circ b_*+b_*\circ \partial)\\
\qquad{} -{(a+\beta_2 \circ \partial)}^2 \circ \partial
 \end{array}
 &
 \begin{array}{l}
(a+\beta_2\partial)(b_*\partial+\partial b_*)\\
\qquad{} +(b_*\partial+\partial b_*)
(a-\beta_2\partial)
  \end{array}
\end{array}
\! \! \! \! \right)}_.
\eeq
\bigskip

It easy to check that $B_*^{II}$ and $B_*^{III}$ are hamiltonian. We have
to take

\beq
\label{hzero}
  H_{*0}=\int\limits_{S^1} b_*(x) \;dx
\eeq

and change arguments ${\breve b}_* \mapsto b_*$ in variational derivatives.

 Equations of (\ref{eulerkuper}) with $B = B^{II}_*$ with $H_{*1}$ describes
a simple dynamical system

\beq
\label{hfirstdinour}
\left\{
\begin{array}{rcl}
{\dot a}  &=&  a'\\
{\dot b_*} &=&  b_*'.\\
\end{array} \right.
\eeq

  Now let us get back to the original problem. Systems (\ref{bstardin}) and
(\ref{euforus}) are equivalent up to conditions $\beta_3 \ne 0, \beta_2
\ne0$. Therefore we may conclude that (\ref{mdww}) is also integrable.
Leaving the first structure ${\breve B}^{I}$ (\ref{kufirststrabreve})
unchanged we may present (\ref{mdww}) in the form of three-hamiltonian
structure equation with hamiltonian matrixes

\bigskip
 \beq
 \label{matrixhatb}
 B={\widehat B}^{II} =
 \left(
 \begin{array}{ccc}
\beta_{1} \circ \partial && \partial
 \circ a - \beta_{2} \circ {\partial}^2 \\
 && \\
a \circ \partial + \beta_{2}
 \circ {\partial}^2 && b \circ \partial + \partial \circ b  +\beta_{3} \circ
{\partial}^3
\end{array} \right)
\eeq
\bigskip


\bigskip

 \beq
\label{thirdhamstra}
{\widehat B}^{III}= \! {\left( \!
\begin{array}{ll}
\beta_1(\partial\circ a+a\circ\partial) &
\begin{array}{l}
  \beta_1 (\partial \circ b+b\circ \partial)\\
  \qquad{} +\partial \circ {(a-\beta_2 \circ \partial)}^2\\
 \qquad{} \qquad{} +\beta_3 {\partial}^3
\end{array}
  \\
 {}&{}\\

\begin{array}{l}
\beta_1(\partial\circ b+b\circ\partial)\\
\qquad{} -{(a+\beta_2\circ\partial)}^2\circ \partial \\
\qquad{} \qquad{} +\beta_3{\partial}^3
\end{array}
 &
 \begin{array}{l}
(a+\beta_2\partial)(b\partial+\partial b)\\
\qquad{} +(b\partial+\partial b)
(a-\beta_2\partial)\\
\qquad{} \qquad{} +\beta_3(a{\partial}^3+{\partial}^3a)
 \end{array}
\end{array}
\! \! \right)}_.
\eeq
\bigskip

  With the help of three-hamiltonian structure we can calculate the whole
family of Hamiltonians for the systems (\ref{mdww}). For example the third
Hamiltonian is

\beq
\label{htilde3}
 H_3=\int_{S^1} \left( a^2b+\beta_2(ab'-a'b)+\frac{1}{2}\beta_1b^2
-\frac{1}{2}\beta_3(a')^2 \right) \;dx
\eeq

while in the first and second coincide with $H_{*1}$ and $H_{*2}$ with
$b_*$ substituted by $b$.

  As it is shown in \cite{kupershmidt1} the DWW system of equations has for
every odd flow $n=1\;(mod \; 2)$ the invariant manifold defined by $a=0$ on
which this hierarchy reduces to the KdV hierarchy.

\bigskip
\bigskip
\bigskip
\bigskip
 Now let us take a standard quadratic Hamiltonian

\beq
\label{hfirst}
{\widehat H}_{1}=\int_{S^1} \frac{1}{2}\left( b^2+a^2\right)\;dx
\eeq

or

\beq
\label{htildefirst}
{\widetilde H}_1=\int_{S^1} \frac{1}{2}\left({\widetilde b}^2+a^2\right)\;dx.
\eeq

Then in (\ref{abm})

\beq
\label{mhfirst}
M^{-1}=M^{-1}_{KdV}=
\left( \begin{array}{cc}
1&0\\
0&1
\end{array}
\right).
\eeq

 We derive $f=b, g=a$ from the Hamiltonian ${\widehat H}_1$ (\ref{hfirst})
and write (\ref{eulerkuper}) with the same $B={\widehat B}^{II}$ as before

\beq
\label{hquafirstdin}
\left\{ \begin{array}{rcl}
{\dot a}&=&(ba+\beta_1a-\beta_2b')'\\
{\dot b}& = & \beta_3b'''+3bb'+aa'+\beta_2a''.
\end{array} \right.
\eeq

 We have $f={\widetilde b}$,$a=g$ for the system (\ref{btildadin}) with
${\widetilde H}_1$

\beq
\label{hbtildedin}
\left\{ \begin{array}{rcl}
{\dot a}&=&({\widetilde b} a+\beta_1a-\beta_2{\widetilde b}')'\\
{\dot {\widetilde b}} & = & \;\gamma {\widetilde b}'''+
3{\widetilde b}\,{\widetilde b}'
\end{array} \right. .
\eeq

  One can see, that the second equation in (\ref{hbtildedin}) is the KdV
equation. The infinite number of its conservation laws allows us to
integrate the first equation of our system.

  Recall that the systems (\ref{euforus}) and (\ref{btildadin}) are
equivalent under conditions $\beta_1\ne0, \beta_2\ne0$. Therefore as in
the case of DWW systems we conclude that (\ref{hquafirstdin}) is integrable.

 As a generalization of bi-hamiltonian structure of the KdV equation the
system (\ref{hquafirstdin}) has a form

\bigskip
\beq
\label{doublestructure}
{a \choose b}_t=
{\widehat B}^{I}
\left(
\begin{array}{c}
 {\frac{\delta {\widehat H}_{k+1}}{\delta a}}\\{}
 \\
  {\frac{\delta {\widehat H}_{k+1}}{\delta b}}
\end{array} \right)=
{\widehat B}^{II}
 \left(
\begin{array}{c}
 {\frac{\delta {\widehat H}_k}{\delta a}}\\{}
  \\
  {\frac{\delta {\widehat H}_k}{\delta b}}
\end{array}
\right)
\eeq
\bigskip

where

\beq
\label{hamstrahfirst}
{\widehat B}^{I}=
\left(
\begin{array}{cc}
\partial &0\\
0 &\partial
\end{array}
\right)
\eeq
 $k=0,1$ and

\beq
\label{h0}
{\widehat H}_0=\int_{S^1}(a(x)+b(x))\;dx.
\eeq

((\ref{eulerkuper}) with ${\widetilde H}_0$ is again (\ref{hfirstdinour}))

\beq
\label{hsecond}
{\widehat H}_2=\int_{S^1} \left( \left[ \frac{1}{2}b^3-\frac{\beta_3}{2}
{\left(\frac{\partial b}{\partial x}\right)}^2 \right]+\left[ \frac{1}{2}
a^2b+\beta_2(a'b-ab')+\beta_1\frac{1}{2}a^2 \right] \right) \;dx.
\eeq

  It easy to see that ${\widehat H}_0$, ${\widehat H}_1$ and ${\widehat
H}_2$ are integrals of motion. Indeed \cite{das}, equations in
(\ref{hfirstdinour}) and (\ref{hquafirstdin}) are in the form of the
continuity equation. If we denote $q_0=a+b$ for ${\widehat H}_0$ and
$q_1=a^2+b^2$ for ${\widehat H}_1$ then

\beq
\label{contequation}
\partial_tq_k+\partial j_k=0
\eeq

$k=0,1$ and $q_0$,$q_1$ are conserved densities. Using the two-hamiltonian
structure we can calculate higher Hamiltonians of the system. Analogous but
less interesting dynamical systems appear from (\ref{bstardin}).

 The system (\ref{hquafirstdin}) can also be transformed in a way that
clarify its coupling with the family of Hamiltonians
(\ref{brevehamiltonians}).  Let

\beq
\label{bhat}
{\widehat b}=b+\beta_1
\eeq

then

\beq
\label{bhatsys}
\left\{
\begin{array}{rcl}
{\dot a}&=&({\widehat b} a-\beta_2{\widehat b}')'\\
{\dot {\widehat b}} & = & \;\beta_3 {\widehat b}'''+
3{\widehat b}\,{\widehat b}'+aa'+\beta_2a''-3\beta_1{\widehat b}.
\end{array}
\right.
\eeq

 (\ref{bhatsys}) has the form of (\ref{doublestructure})
with the same ${\widehat H}_0$, ${\widehat H}_1$ ,${\widehat B}^{I}$ ($b$
is replaced by ${\widehat b}$) and

\bigskip
\beq
\label{matrixbhat}
  {\widehat B^{II}}_{\widehat b} =
  {\left(
  \begin{array}{ccc}
 0 && \partial
 \circ a - \beta_{2} \circ {\partial}^2 \\
 && \\
a \circ \partial + \beta_{2}
 \circ {\partial}^2 && {\widehat b} \circ \partial + \partial
 \circ {\widehat b}  +\beta_{3} \circ
{\partial}^3 -3\beta_1 \circ \partial
\end{array} \right)}_.
\eeq
\bigskip

The expression in the second square brackets reminds us ${\breve H}_3$

\beq
\label{bhathamsecond}
{\widehat {\widehat H}}_2=\int_{S^1} \left( \left[ \frac{1}{2}{\widehat
b}^3-\frac{\beta_3}{2}{\left(\frac{\partial {\widehat b}}{\partial
 x}\right)}^2-3\beta_1{\widehat b}'\right] +\left[\frac{1}{2}a^2b+\beta_2
(a'b-ab') \right] \right) \;dx
\eeq

 It is interesting to mention that the condition $b=0$ in ${\widetilde b}$
leads to the reduction of (\ref{hbtildedin}) to

\beq
\label{kdvmkdv}
\left\{
\begin{array}{rcl}
\bigskip
{\dot a}&=&\beta_1a'+\frac{1}{\beta_1}(\beta^2a'''-\frac{3}{2}a^2a')
\\ {\dot {\widehat a}} & = & \;\gamma {\widehat a}'''+ 3{\widehat a}\,
{\widehat a}'
\end{array} \right.
\eeq

with

\beq
\label{widehata}
{\widehat a}=-\frac{\beta_2}{\beta_1}a'-\frac{1}{2\beta_1}a^2
\eeq

so that we have a mKdV equation for $a$ and a KdV equation for ${\widehat
a}$ related by some kind of Miura transformation.

The system (\ref{bstardin}) with $f=b_*$ and $g=a$ gives

\beq
\label{hstardin}
\left\{ \begin{array}{rcl}
{\dot a}&=&(b_*a+\beta_1a-\beta_2b_*')' \\
{\dot b_*} & = & 3b_*b_*'+aa'+\beta_2a''.
\end{array} \right.
\eeq

 As it was in (\ref{hbtildedin}) for $b=0$ we obtain a pare of mKdV and KdV
equations when $b=0$

\beq
\label{mkdvkdv*}
\left\{
\begin{array}{rcl}
\bigskip
{\dot a}&=&\frac{\beta_3} {{\widetilde \gamma} } ( \frac{3} {2{\beta_2}^2}
a^2a'-a''')-\beta_1a' \\
 {\dot {\widehat a}_*} & = & \;\gamma {\widehat a}_*'''+ 3{\widehat a}_*\,
{\widehat a}_*'
\end{array} \right.
\eeq

where

\beq
\label{gammatilde}
{\widetilde \gamma}=1+\frac{\beta_3\beta_1}{{\beta_2}^2}
\eeq

\beq
\label{hatast}
{\widehat a}_*=\frac{1}{{\widetilde \gamma}}(\frac{\beta_3}{\beta_2}a'+
\frac{\beta_3}{2{\beta_2}^2}a^2).
\eeq

 The Lax representation for the systems (\ref{euforus}) and (\ref{bstardin})
with two types of Hamiltonians is obvious (\ref{laxform})

\beq
\label{laxformt}
\partial_t L= {ad}^*_V L
\eeq
and
\beq
\label{laxformstart}
\partial_t L_*= {ad}^*_V L_*
\eeq
\bigskip

(see (\ref{diffoper}), (\ref{amatrix}), (\ref{ourL}) and
(\ref{ourLstar})) where elements of the algebra should be replaced by
coalgebra elements.


  (\ref{euforus}) (with $\beta_1=\beta_2=\beta_3=0$ ) can also be considered
as a system of equation of motion for generalized magnetic hydrodynamics in
the configuration space $G$, \cite{vishik}

\beq
\label{magnitodinamics}
\begin{array}{c}
\bigskip
{\dot {\cal M}}=-ad^*_{\cal N} {\cal M}+ad_{\cal H}^*J\\
{\dot {\cal H}}=\left[ {\cal N},{\cal H} \right]
\end{array}
\eeq

 where the second equation can be written in the $ad^*$-terms in coalgebra
so that ${\cal N}=-f,{\cal H}=g,{\cal M}=b,J=a$ if $(q,J) \in {\cal G}^*$
corresponding to $(-f,g) \in {\cal G}$.

 It might be interesting to consider small oscillations in the dynamical
system described by (\ref{hquafirstdin}). We find

\beq
\label{smallascill1}
-i(\omega -\beta_3k^3)b=-i \beta_2k^2a
\eeq

\beq
\label{smallascill2}
-i\omega=\beta_2k^2
\eeq

which leads us to

\beq
\label{omegak}
\omega=\frac{\beta_3k^3}{2} \pm \sqrt{\frac{{\beta_3}^2k^6}{4}+\beta_2k^4}
\eeq

as a $\omega(k)$-dependence.


\section{Commuting flows and Poisson brackets on~${\widehat {\cal {G}}}^{*}$ }
\qquad {} There exists a standard way to obtain commuting
 flows on coadjoint orbits of $G$, \cite{segal1} . For two
elements $h_1,h_2 \in {\widehat {\cal G}}$ and $u \in {\widehat {\cal G}}
^{*}$ the Poisson bracket is given by

\beq
\label{poissonbracket}
\left\{ h_1,h_2 \right\}(u)=\langle u,\left[ dh_1,dh_2 \right] \rangle.
\eeq

 Choose a fixed element ${\bf q}=(q_1,q_2,...) \in {\widehat {\cal
G}}^{*}$.  Then for each $G$-invariant function $F$ on ${\widehat {\cal
G}}^{*}$, (i.e.

\beq
\label{ginvariance}
\langle \tau .h,dF(h) \rangle =\langle h,\left[ \tau,dF(h) \right] \rangle
=0,
\eeq

for every $\tau \in {\widehat {\cal G}}$ and every $h \in \widehat {\cal
G}^{*}$) and for each $\chi=(\chi_i),\; \chi_i \in {\bf C}$ functions

\beq
\label{invarfunction}
F_{\chi}(h)=(F_{\chi,1}(u_1-\chi_1 q_1),\;F_{\chi,2}(u_2-\chi_2 q_2),...)
\eeq

Poisson-commute \cite{segal1}. The proof is a simple generalization of the
lemma in \cite{segal1} using the skew symmetry of the Poisson bracket
(\ref{poissonbracket}) of multi-component functions (cf. \cite{segal1}).

  Consider now an affine Lie algebra $ {\cal L} ({\cal G})=\sum\limits_{i
< 0}^{} \varsigma_i \lambda^i + \sum\limits_{i \ge 0}{}
 \varsigma_i \lambda^i $
consisting of Lorant polynomials  with
coefficients $ {\varsigma }_{i} \in \cal G $. Let $ \widehat {\cal L }
 (\widehat {\cal G }) $ be the loop algebra connected to $\widehat {\cal G }
$ and centrally extended by ${R}{\! \! \! \! \! \,}{R}\; (  \lambda ,
{\lambda}^{-1} ) $. Let $R=P_{+}-P_{-}$ be a standard $r$-matrix where
$P_{\pm}$ are projection operators from $\cal L$ to corresponding
subalgebras.  The dual to $\widehat {\cal {L} }(\widehat {\cal G})$ space
is introduced by means of the pairing (see \cite{semenov1},\cite{semenov2})

\beq
\label{looppairing}
\begin{array}{l}\bigskip
 \langle (b,a,\beta_1,\beta_2,\beta_3),(f,g,\alpha_1,\alpha_2,\alpha_3)
\rangle=\\
\qquad{} \qquad{} {{Res}_{\lambda=0}} \left( \int\limits_{} (b\;f
+a\;g)\;dx
+\sum\limits_{i=1}^{3} \beta_i\alpha_i \right).
\end{array}
\eeq

In the space of polynomials of order $\le N$ (with fixed $\beta_m,\;m=1,2,3$)

\beq
\label{newpoisson}
\chi(\lambda,x)=
{ {a(\lambda,x)} \choose {b(\lambda,x)} }_{\beta_1,\beta_2,\beta_3}
= \sum\limits_{i=0}^{N}\chi_i(x)\lambda^i
\eeq

 we have three linear systems of $N+2$ compatible brackets. Two first of
them (compare with (\ref{euforus})) are

\beq
\label{newfirstbracket}
\bigskip
 {\left\{ \chi_i(x),\chi_j(y) \right\} }_
{R {\widehat {\lambda}}^{k} }^{DWW_1}=
\epsilon \sum\limits_{s=0}^{N} \!
\left(
\begin{array}{cc}
0&\partial\\
\partial & 0
\end{array}
\right)
\left(
\! \!  \begin{array}{c}
\delta(x-y)\\
{}\\
\delta(x-y)
\end{array}
\! \right)
\eeq

\beq
\label{newbrackets}
\begin{array}{l}
\bigskip {\left\{ \chi_i(x),\chi_j(y) \right\} }_
{R {\widehat {\lambda} }^{k}}^{DWW_2}=\\
\epsilon \sum\limits_{s=0}^{N} \!
\left(
\! \begin{array}{ll}
\beta_{1,s}{\delta}_{l,s}\partial &
a_l(x)\circ\partial-
\beta_{2,s}{\delta}_{l,s}{\partial}^2\\
{}&{}\\
a_l(y)\circ\partial+
\beta_{2,s}{\delta}_{l,s}{\partial}^2 &
\begin{array}{l}
b_{l}(x)\circ\partial+b_{l}(y)\circ\partial\\
 \qquad{} \qquad{} +\beta_{3,s}
 {\delta}_{l,s}{\partial}^3
 \end{array}
\end{array}
\! \! \! \right)
\! \! \left(
\! \!  \begin{array}{c}
\delta(x-y)\\
{}\\
\delta(x-y)
\end{array}
\! \right)
\end{array}
\eeq
\bigskip

where $l=i+j+1-k, \;k=0,...,N+1,\; \epsilon=1$ when
$i,j \ge k$,$\epsilon=-1$ when $i,j<k$,$\epsilon=0$
for $i \ge k$,$j<k$.

 The derivation of the third system  of brackets from (\ref{thirdhamstra}) is
rather tedious but simple exercise.

  In the KdV-like case the system of second brackets coincides with
(\ref{newbrackets}) while the first is

\beq
\label{newfirstbracketkdv}
\bigskip {\left\{ \chi_i(x),\chi_j(y) \right\} }_
{R {\widehat {\lambda}}^{k} }^{KdV_1}=
\epsilon \sum\limits_{s=0}^{N} \!
\left(
\begin{array}{cc}
\partial&0\\
0 & \partial
\end{array}
\right)
\left(
\! \!  \begin{array}{c}
\delta(x-y)\\
{}\\
\delta(x-y)
\end{array}
\! \right)
\eeq

  As for the third Poisson bracket for (\ref{hquafirstdin}) one could try to
generalize the non-local bracket found in \cite{rubtsorlo} for our system of
equation.


\section{Conclusion}

 To complete the analysis of hamiltonian strictures considered in this
 paper, it would be interesting to try to find a whole family of commuting
Hamiltonians for the KdV-like system (\ref{euforus}) using the
bi-hamiltonian form of it (\ref{doublestructure}) or the resolvent of the
inverse to the Lax operator (\ref{ourL}) ( cf. \cite{segal1}).  It
is also possible to write alternative Lax representation for our systems
using the third hamiltonian structures.


\section{Acknowlegments}
    The author is deeply indebted to his supervisors M.A. Olshanetsky and
V.N. Rubtsov for pointing out the problem and stimulating discussions. He
would also thank A. Gorsky, E.A. Kuznetsov, D. Ivanov, A. Blinov and
A. Antonov for very useful suggestions and comments.

\end{document}